\documentstyle[12pt]{article}

\input{tcilatex}
\begin{document}

\title{{\bf Edge Modes Waves in Superlattices in Quantum Hall Effect Regime}}
\date{}
\author{}
\maketitle

\begin{center}
\vspace{-1.0cm} {\large {\bf Pavel Fileviez Perez,}} \\[0pt]
\vspace{.25cm} {\small {\it Instituto Superior de Ciencias y Tecnologia
Nucleares}} \\[0pt]
{\small {\it Salvador Allende y Luaces, Vedado, La Habana, Cuba,}}\\[0pt]
\vspace{.5cm} {\large {\bf Alejandro Cabo Montes de Oca}}\\[0pt]
\vspace{.25cm} {\small {\it Center of Theoretical Studies of Physical Systems%
}}\\[0pt]
{\small {\it Clark Atlanta University, Atlanta, Georgia, U.S.A.}}\\[0pt]
{\small {\it and }}\\[0pt]
{\small {\it Instituto de Cibernetica Matematica y Fisica}}\\[0pt]
{\small {\it Calle E, N. 309, Esq. a 15, Vedado, La Habana, Cuba}}\\[0pt]
\vspace{.5cm} 
{\large {\bf Carlos Rodriguez Castellanos}} \\[0pt]
\vspace{.25cm} {\small {\it Grupo de Fisica Teorica, Facultad de Fisica,
Universidad de la Habana}} \\[0pt]
{\small {\it San Lazaro y L, Ciudad de la Habana, Cuba}}\\[0pt]
\vspace{1.0cm} {\large {\bf Abstract}}\\[0pt]
\end{center}

\noindent
The wave propagation of edge modes in a superlattice of 2D electron Gases in
quantum Hall regime is investigated. After introducing surfaces charge and
current densities at the edge, the Maxwell equations are solved for waves
running along the boundary. The constitutive relations expressing the edge
charge and current densities in terms of the fields at the boundary are
derived. One of them is similar to the London equation for superconductor
currents. The dispersion relation and wave polarizations for the momenta
region $w/c<k$ are also obtained for propagation along the borderlines of
the electron gases. It follows that the modes have no dispersion at any
frequency. The static limit solutions complete the definition at the
boundary of the formerly determined interior field configurations showing a
Meissner like effect. The results underline that various of the current
theoretical approaches to edge excitations could be appropriate for
superlattice structures but can fail to describe standard planar samples. 
\newpage 

This work is devoted to investigate the propagation properties of edge modes
in superlattices formed by 2D electron gases under quantum Hall regime [1].
Such structures are constituted by independent semi-planes in which the
electrons are mobile only within their semi-plane. The distance between the
planar gases will be assumed to be small with respect to the wavelengths
under investigation. Then the system can be considered as an anisotropic
medium. The typical width of the 2D electron gases is the order of $10-100\
A^0$ and the distance between them can be of the order of hundreds or
thousands $A^0$ in normal situations. The physical properties of the
excitations at the border of samples showing fractional or integral QHE have
been the objective of study from early times after the discovery of the
effect [2-9]. The motivation for their study got further strength after the
development of the Buttiker description [3]. One of the points of central
interest was the study of the electromagnetic properties near the edges
[10]-[19]. In spite of the extensive attention to this problem in the
literature, in our view, there remain yet important unsolved questions.
Among them, stays the structure of the edge field configurations in planar
samples. To be specific, the dynamics of current and charges at the border
furnishing the gauge invariance (or charge conservation) of the action
describing the QHE(Chern-Simons action) [10] are not yet fully understood
for realistic planar samples. We think that the main difficulty in the
development of such an understanding is related with the assumption about
the 2 space and one time nature of the electromagnetic action. It is certain
that the Chern-Simons contribution has effectively this structure due to the
planar character of the currents [12]. However, the electromagnetic action
should retain its usual 3-space +1 time structure because this form accounts
for the real electromagnetic field to be non planar for 2D samples [12,17].
A justification of these statements was done in Refs. [12,17]. The effective
Maxwell equations derived in that work were investigated in Refs. [16,18].
These equations generalized previous ones valid for the static limit
[4,7,8]. Their study furnished interesting conclusions as the existence of:
Meissner effect in superlattices [2] and surface waves in planar samples
[18]. Further they also allowed to propose a natural model for a surprising
tunneling phenomena [20] in strip samples in QHE regime [16]. Therefore,
when superlattice structures of planar QHE samples are considered [12] the
field distribution can be two dimensional, that is, invariant along the
normal directions to the planes. However, the pure 2D samples seem to need
for a more involved analysis mainly due to the singular behavior of strictly
filamentary edge charge and current densities [18]. The analysis of this
problem will be considered elsewhere.

The objective of the present work is to investigate the propagation
properties of edge mode waves in superlattice of planar samples. It should
be stressed that the discussion done in Ref. [10] was restricted by the very
low momentum approximation. \smallskip The interpretation for the equations
to be valid for superlattices also introduce the possibility of studying the
propagation of modes in any direction ranging from the parallel to the
normal to the planar gases. After introducing surface charge and current
densities at the edge, the Maxwell equations are solved for waves
propagating within the boundary. The constitutive relations expressing the
densities in terms of the field values at the boundaries are determined.
Their form resembles the London equation for the superconductor currents. In
the momenta region $w/c<k$ the dispersion and polarization vectors have been
derived. The propagation along the direction parallel to the planes turns to
be non dispersive at any momenta, then verifying the assumption posed in
[10].

It can be useful to underline again that our discussion is closely linked
with that given by Wen [10] which in common with other authors have pointed
out the relevance of the Chern-Simons action for the description of the
electromagnetic properties of QHE samples. However, as it was indicated in
[12], the treatment in [10] turns to be appropriate only for strictly 2+1
dimensional systems. That is, it should become applicable for superlattice
arrays of 2D electronic gases. One of the objectives of this work is to
underline this circumstance.

The analysis here presented can be considered as complementing the results
of [10] by determining the structure of the electromagnetic propagation
outside the infrared non dispersive region considered in that reference.
Furthermore, the results for the static field configurations reproduce the
ones obtained in [12] which were used to argue the existence of a Meissner
like effect in superlattices [12]. Here the solution is fully determined by
finding the charge and current at the border in terms of the dynamical
fields and the outside field values. As it was mentioned above, the current
satisfies a sort of London relation.

It should be stressed that these particular static solutions can be used to
investigate the physical nature of the transitions between different
plateaus at varying magnetic fields. Such a study would be one of the
possible extensions of the present work.

Another worth to be addressed task can be the study along the same lines
followed here of single planar samples. This task would have as a
consequence a better understanding of the transient propagation at the
border of realistic planar samples which has been the subject of
experimental examination [20].

\section{The Maxwell equations.}

Let us consider the Maxwell equations for superlattices of 2D electron gases
in QHE regime discussed in [12]:

\begin{equation}
\Box A_\mu \left( r,x_4\right) =-\frac{4\pi }cJ_\mu \left( r,x_4\right) ,\ 
\end{equation}

\[
\partial _\mu A_\mu =0, 
\]

\noindent where the current density has two contributions, the volumetric
currents associated with the array of independent 2D electron gases and the
surface sources at the border which should be determined in order to satisfy
the Maxwell equations, boundary conditions and charge conservation. Thus, we
write

\begin{equation}
J_\mu \left( r,x_4\right) =J_\mu ^{sp}\left( r,x_4\right) +J_\mu
^{edge}\left( r,x_4\right)
\end{equation}

\noindent where the volumetric current in terms of the vector potential has
the following form [12]

\[
J_\mu ^{sp}\left( r,x_4\right) =i\sigma _HI_0\Theta \left( x_1\right)
\varepsilon ^{\alpha \mu \sigma \nu }n_\alpha \frac{\partial A_\mu }{%
\partial x_\sigma }+cI_0\Theta \left( x_1\right) \chi _e* 
\]

\begin{equation}
\ast \left[ P_{\mu \nu }u_\alpha ^2\frac{\partial ^2A_\nu }{\partial
x_\alpha ^2}-u_\mu P_{\nu \alpha }u_\beta \frac{\partial ^2A_\nu }{\partial
x_\alpha \partial x_\beta }-u_\mu P_{\mu \alpha }u_\beta \frac{\partial
^2A_\nu }{\partial x_\alpha \partial x_\beta }+u_\mu u_\nu P_{\alpha \beta }%
\frac{\partial ^2A_\nu }{\partial x_\alpha \partial x_\beta }\right]
\end{equation}
.

\noindent 
In the above equation $\sigma _H$ is the Hall conductivity, $I_0$ is the
longitudinal density of semi-planes in the $x_3$ direction, $\chi _e\ $is
the planar dielectric constant which occurs to be inversely dependent of the
magnetic field. The sources at the border are written as:

\begin{equation}
J_\mu ^{edge}\left( r,x_4\right) =I_0j_\mu \delta \left( x_1\right)
\end{equation}

\noindent and the various 4-vectors and the tensor involved in (3) are
defined by

\begin{equation}
j_\mu =\left( 0,j_{edge},0,ic\rho _{edge}\right) ,
\end{equation}

\begin{equation}
u_\mu =\left( 0,0,0,1\right) ,
\end{equation}

\begin{equation}
n_\mu =\left( 0,0,1,0\right) ,
\end{equation}

\ \hspace{2in}$P_{\mu \nu }$\ =$\left( 
\begin{array}{llll}
1 & 0 & 0 & 0 \\ 
0 & 1 & 0 & 0 \\ 
0 & 0 & 0 & 0 \\ 
0 & 0 & 0 & 0
\end{array}
\right) ,$

\noindent where the auxiliary charge $\rho _{edge}$ and the current $%
j_{edge} $ densities are introduced. The also auxiliary 4-vectors $u_\mu $
and $n_\mu $ correspond respectively to the 4-velocity of the sample and a
4-vector which is orthogonal to the planar electron gases forming the
superlattice. $P_{\mu \nu }$ is the projector tensor on the space orthogonal
to $n_\mu $. These quantities were introduced in Ref. [12]. Such covariant
entities were useful there for clarifying how the Chern-Simons topological
action is involved in the description of the QHE phenomena.

In starting, let us search for traveling waves of the following form

\begin{equation}
A_\mu =F_\mu e^{ik_1x_1+ik_2x+wx_4/c},
\end{equation}

\begin{equation}
j_{edge}=i_0e^{ik_2x+wx_4/c},
\end{equation}

\begin{equation}
\rho _{edge}=\rho _0e^{ik_2x+wx_4/c}
\end{equation}

\noindent  which already assumes the harmonic character in the direction of
the $x_2\ $axis. The $x_3$ axes is orthogonal to the electron gases.

Here we will consider the situation in which only the Hall conductivity
response of the medium is retained. This assumption corresponds to a limit
of high magnetic fields when $\chi _e=0$ because the effective dielectric
constant is inversely proportional to the magnetic field. Accordingly, the
Maxwell equations for internal region are reduced to:

\begin{equation}
\left( k^2\delta _{\mu \nu }-\frac{4\pi }c\sigma _HI_0\varepsilon ^{\alpha
\mu \sigma \nu }n_\alpha k_\sigma \right) F_\nu =0
\end{equation}

\noindent where the 4-vector $k$ entering in this relation is defined by

\begin{equation}
k_\sigma =\left( 
\begin{array}{c}
k_1 \\ 
k_2 \\ 
0 \\ 
\frac w{ic}
\end{array}
\right)
\end{equation}

At the exterior the same expression will be valid after fixing $\sigma _H=0$.

Equations (11) have two types of volumetric waves satisfying the dispersion
relations $k^2=0\ $and $\ k^2+\sigma ^2=0$\ where 
\[
\sigma =\frac{4\pi }c\sigma _HI_0. 
\]

The polarization vectors of both modes can be written in the common form

\begin{equation}
F_\mu =\left( 
\begin{array}{c}
\frac{\left( k_4k_1-k_2\sigma \right) }{k_4^2+\sigma ^2}F_4 \\ 
\frac{\left( k_2k_4+k_1\sigma \right) }{k_4^2+\sigma ^2}F_4 \\ 
F_3 \\ 
F_4
\end{array}
\right) .
\end{equation}

\noindent
where $F_3$ is independent of $F_4$. Now we will consider the two spatial
regions of the problem: the free space at $\ x_1<0$ and the superlattice
bulk $x_1>0.$

\subsection{Region $x_1<0$}

In this zone the waves always satisfy \ $k^2=0\ $and then the vector
potential takes the explicit form

\begin{equation}
A_\mu ^{ext}=\left( 
\begin{array}{c}
A_1 \\ 
A_2 \\ 
A_3 \\ 
A_4
\end{array}
\right) e^{\lambda x_1+ik_2x+wx_4/c}
\end{equation}

\noindent where the real constant

\begin{equation}
\lambda =\sqrt{k_2^2-\frac{w^2}{c^2}}
\end{equation}

\noindent defines a decaying character of the fields at $x_1\rightarrow
-\infty $.

After also imposing the Lorentz gauge condition$\ \partial _\mu A_\mu =0,$
the following relation arises :

\begin{equation}
\lambda A_1+ik_2A_2+\frac wcA_4=0.
\end{equation}

\subsection{Region $x_1>0$}

In this zone two possible waves exist. Those obeying $k^2=0\ $and having a
vector potential in the following form:

\begin{equation}
A_\mu ^{(2)}=\left( 
\begin{array}{c}
\frac{q_2}{k_4} \\ 
\frac{k_2}{k_4} \\ 
A_3/C_1 \\ 
1
\end{array}
\right) C_1e^{-\lambda x_1+ik_2x+wx_4/c},q_2=i\lambda ,
\end{equation}

\noindent and modes satisfying $k^2+\sigma ^2=0\ $and having polarizations

\begin{equation}
A_\mu ^{(1)}=\left( 
\begin{array}{c}
\frac{(k_4q_1-k_2\sigma )}{k_4^2+\sigma ^2} \\ 
\frac{(k_2k_4+\sigma q_1)}{k_4^2+\sigma ^2} \\ 
0 \\ 
1
\end{array}
\right) C_2e^{-\lambda _1x_1+ik_2x+wx_4/c}
\end{equation}

\noindent where

\begin{equation}
\lambda _1=\frac{q_1}i=\sqrt{k_2^2-\frac{w^2}{c^2}+\sigma ^2}
\end{equation}

\noindent which again defines a decaying behavior now at $\ x_1\rightarrow
\infty $.

Systematically using the boundary conditions for the electromagnetic
potentials, the following relations arise

\begin{equation}
\lambda A_{_4}+\lambda C_1+\lambda _1C_2=4\pi iI_0\rho _0,
\end{equation}

\begin{equation}
\lambda A_2+\frac{k_2}{k_4}C_1\lambda +\frac{(k_2k_4+\sigma q_1)}{%
k_4^2+\sigma ^2}C_2\lambda _1=\frac{4\pi }cI_0i_0,
\end{equation}

\begin{equation}
A_1=\frac{q_2}{k_4}C_1+\frac{(k_4q_1-k_2\sigma )}{k_4^2+\sigma ^2}C_2,
\end{equation}

\begin{equation}
A_2=\frac{k_2}{k_4}C_1+\frac{(k_4k_2+q_1\sigma )}{k_4^2+\sigma ^2}C_2,
\end{equation}

\begin{equation}
A_4=C_1+C_2,
\end{equation}

\begin{equation}
A_3=0.
\end{equation}

Further, after imposing the charge conservation at the boundary, which
relates the edge charge or current density and the Hall volumetric currents,
it follows the additional condition :

\begin{equation}
i\sigma _H\left[ ik_2\left( C_1+C_2\right) -\frac wc\left( \frac{q_2}{k_4}%
C_1+\frac{(k_4k_2+q_1\sigma )}{k_4^2+\sigma ^2}C_2\right) \right]
+ik_2i_0+iw\rho _0=0.
\end{equation}

Let us consider first the static situation corresponding to $k_2=0,k_4=0.$
Then, from (20)-(26) it is possible to find the following simple relations:

\begin{equation}
A_4=C_2=\frac{ic\rho _0}{\sigma _H}=-i\frac{i_0}{\sigma _H}=-i A_2
\end{equation}

\noindent which closely resemble the London equations for superconductivity
where a current is directed along the direction of the spatial vector
potential. There exist also a connection between $\rho _0$ and $i_0$ which
turns to be very simple in this case:

\begin{equation}
i_0=-c\rho _0
\end{equation}

The spatial dependence of the fields in this static regime is simply an
exponential decay which expresses the occurrence of a Meissner like effect
in the superlattice. Such a phenomenon is absent in planar samples as it was
discussed in [12].

Let us return now to the general discussion. It is possible to obtain the
general dispersion relation for propagating modes as the condition for the
existence of nontrivial solutions of the system of linear equations( 16,
20-26). The dispersion relation coming from the vanishing of the relevant
determinant takes the form:

\[
w-\sigma _HI_0\left( k_2+i\frac wcB\right) \frac{4\pi }{\left( \lambda
+\lambda _1-2\lambda \frac{\eta _2}{\eta _1}\right) }+ 
\]

\begin{equation}
+ik_2c\left[ \left( \lambda +\lambda _1\right) B-2\lambda \frac{k_2}{k_4}%
\frac{\eta _2}{\eta _1}\right] /\left( \lambda +\lambda _1-2\lambda \frac{%
\eta _2}{\eta _1}\right) =0
\end{equation}

\noindent in which the new quantities appearing are defined by

\begin{equation}
\frac{\eta _2}{\eta _1}=\frac{\lambda A+ik_2B+w/c}{i\frac{\lambda ^2}{k_4}+i%
\frac{k_2^2}{k_4}+\frac wc},
\end{equation}

\begin{equation}
A=\frac{k_4q_1-k_2\sigma }{k_4^2+\sigma ^2},
\end{equation}

\begin{equation}
B=\frac{k_2k_4+q_1\sigma }{k_4^2+\sigma ^2}.
\end{equation}

In spite of its complicated appearance, the dispersion relation
corresponding to the propagation parallel to the electron planes , have the
simple non dispersive solution

\[
w=c\ k_2 
\]
as is it predicted in [10]. Then, the present results indicate that the
propagation is non dispersive even at high values of the momentum. This
solution for the dispersion relation is common for two modes. One of them is
a wave with the electric fields polarized in the direction perpendicular to
the electron gas planes. This orientation of the electric field produces
that the wave does not interact with the superlattice because the field can
not move the electron out of the planes. Also, the magnetic field produces a
Lorenz force which is also perpendicular to the same planes and also does
not polarize the charges.

The other mode is simple the wave propagating variant of the static edge
modes associated to the quantum Hall effect (or Meissner like) currents
circulating though the borders. The waves decay exponentially at the
interior of the superlattice and are completely undamped at free space. The
propagation properties for general directions of the momenta after the
inclusion of the dielectric properties of the background medium will be
discussed in more detail elsewhere.

\section{Conclusions}

The propagation properties for edge modes in a superlattices of 2D electron
gases in QHE regime is determined. The field dependence of the edge charge
and current densities which implement charge conservation also follows[7].
The solutions for the static case at zero frequency are extending the ones
previously obtained in Ref. [12] to be also valid at the boundary and
external region. Thus, this work fully support the global (in space)
validity of the Meissner effect solutions derived in [12]. As a by-product
the the edge and current densities at the boundary are also determined. The
constant slope of dispersion relation at any momenta indicates that the
waves have are non dispersive.

The results can be generalized for the inclusion of the dielectric
properties of the background media. This would be necessary in order to
compare with possible future experiments. A detailed study of the
propagation along the directions non parallel to the planes will be
considered elsewhere. From the technical point of view the solution of the
here considered task will be also helpful for the investigation of the more
relevant case of planar samples which are the ones employed in most of the
experiments.

\section{Acknowledgments}

This work was supported by the Christopher Reynolds Foundation. One of the
authors (A.C.M) want also to acknowledge the kind hospitality of the CTSPS
at the Clark Atlanta University where the work was completed. He is also
indebted to Dr. C. Handy, Dr. D. Bessis, Dr. X.Q. Wang and Dr. G. Japaridze
for their valuable comments and kind support.


\newpage

\end{document}